\documentstyle[floats,aps,preprint]{revtex}
\begin{document}
\draft
\preprint{\vbox{ \hbox{SOGANG-HEP 264/99} \hbox{hep-th/9911010}  }}
\title{Massless {\it vs}. Massive Hawking Radiation in AdS$_2$ Spacetime}
\author{Sung-Won Kim$^{a}$\footnote{electronic address:sungwon@mm.ewha.ac.kr}, 
        Won Tae Kim$^{b}$\footnote{electronic address:wtkim@ccs.sogang.ac.kr},
        and John J. Oh$^{b}$\footnote{electronic address:john5@gravity.sogang.ac.kr}}
\address{${}^{a}$ Department of Science Education,\\
         Ewha Women's University, Seoul 120-750, Korea\\
         ${}^{b}$ Department of Physics and Basic Science Research Institute,\\
         Sogang University, C.P.O. Box 1142, Seoul 100-611, Korea}
\date{\today}
\maketitle
\bigskip 
\begin{abstract}
We study massless and massive Hawking radiations on 
a two-dimensional AdS spacetime. For the massless case,
the quantum stress-energy tensor of a massless scalar field
on the AdS background is calculated, and the expected null 
radiation is obtained.  However, for the massive case, the
scattering analysis is performed in order to calculate 
the absorption and reflection coefficients which are related
to statistical Hawking temperature. On the contrary to the 
massless case, we obtain a nonvanishing massive radiation.
\end{abstract}
\bigskip
\newpage


There has been a great interest in a lower-dimensional 
gravity since it is possible to construct a consistent and
renormalizable quantum gravity without encountering some complexities of
four-dimensional realistic models. 
In the Callan-Giddings-Harvey-Strominger(CGHS) model 
\cite{cghs}, the asymptotically 
flat black hole solution is obtained 
under a linear dilaton background and the quantum effect of the black hole can be
described by Hawking radiation \cite{haw}. 
In general, a thermal equilibrium state 
of the black hole with the thermal bath is defined by 
the Hartle-Hawking(HH) vacuum \cite{isr,hh}.

>From the CGHS model, 
the two-dimensional anti-de Sitter(AdS$_2$) 
solution can be obtained by assuming 
a constant dilaton background \cite{kim}, which is
in contrasted with the original CGHS solution in that
the curvature scalar of AdS spacetime 
is constant and its asymptotic
metric is no more Minkowski spacetime.
This phenomenon has already appeared in the three-dimensional
low energy string theory \cite{hw},  
so that the asymptotically
flat black string solution is obtained on the
presence of the dilaton charge while the
asymptotically nonflat
Ba$\tilde{{\rm n}}$ados-Teitelboim-Zanelli(BTZ)\cite{btz} 
solution is derived for the constant dilaton background.
On the other hand, in connection with the calculation of the statistical
entropy of the extremal Reissener-Nordstrom
black hole, the AdS$_2$ geometry 
has been intensively studied in 
Refs. \cite{str,ss}. 

On the other hand, it has been shown that
the massless Hawking radiation on this AdS background
does not appear \cite{kim,kop} whereas
it is possible to radiate if the dilaton field 
couples to the massless scalar field \cite{ko}.
It is now natural to ask what happens for the massive case on
this metric background. So we would like to study
whether the massive radiation on the AdS$_2$ spacetime background
is possible or not in terms of the scattering analysis, which
was given for the massless case in Ref \cite{kop}.
In this paper, for the massless case, the stress-energy tensor
calculation will be done on the AdS$_2$ background and the null radiation
will be found, 
which is essentially compatible with the result of the null radiation 
through the scattering analysis \cite{ko}.
In this paper we mainly 
study the massive scalar field on the AdS$_2$ background
and calculate the scattering amplitudes in order to obtain
the Hawking temperature. 
We finally obtain some unexpected result on the Hawking
temperature. 

We first study the massless radiation on the AdS$_2$ background
by calculating the stress-energy tensors of the conformal
matter field. This confirms the expected vanishing Hawking 
radiation. 
Now the two-dimensional action for the conformal matter is given as 
\begin{equation}
  \label{mataction}
  S_{M} = - \frac{1}{2} \int dx^2 \sqrt{-g} \left(\nabla f\right)^2,
\end{equation}
where $f$ is a massless scalar field.
The effective action of the scalar field is 
written as the Polyakov induced gravity action,
\begin{equation}
\label{effectaction}
S_{eff} = - \frac{1}{96\pi} \int dx^2 \sqrt{-g}R\frac{1}{\Box}R,
\end{equation}
and equally written as a local form of
\begin{equation}
  \label{localeffaction}
  S_{eff} = -\frac{1}{96\pi} \int dx^2 \sqrt{-g}[ - \Psi \Box \Psi + 2
  \Psi R],
\end{equation}
by introducing the auxiliary field $\Psi$ satisfying 
\begin{equation}
\label{auxiliary}
\Box \Psi = R.   
\end{equation}
The stress-energy tensor is 
\begin{eqnarray}
  \label{stresstensor}
  <T_{\mu \nu}> &=& \frac{2 \pi}{\sqrt{-g}} \frac{\delta S_{eff}}{\delta
  g^{\mu \nu}} \nonumber \\ &=& -\frac{1}{48} \left[ 2 \nabla_{\mu}
  \nabla_{\nu} \Psi - \nabla_{\mu}\Psi \nabla_{\nu}\Psi - g_{\mu \nu}
  \left( 2 R - \frac{1}{2} (\nabla \Psi)^2 \right)\right]
\end{eqnarray}
where the background metric may be assumed to be in the form of
$ds^2=-g(r)dt^2 + \frac{1}{g(r)}dr^2$.
In the light-cone coordinates, they are explicitly rewritten as
\begin{eqnarray}
  \label{relations}
  <T_{\pm \pm}(\sigma^{+}, \sigma^{-})> &=& \frac{1}{4}\left( <T_{tt}> + g^2(r) <T_{rr}> \pm
  2g(r)<T_{rt}>\right), \\
   <T_{+-}(\sigma^{+}, \sigma^{-})> &=& \frac{1}{4}\left( <T_{tt}> - g^2(r) <T_{rr}>\right)
\end{eqnarray}
where $\sigma^{\pm} = t \pm r^{*}$ and 
$r^{*}(r) = \int dr \frac{1}{g(r)}$.

To derive the Hartle-Hawking temperature for 
the massless scalar
field on the two dimensional AdS$_2$ background,
we consider the AdS black hole metric  given as
a solution of the CGHS model with the constant 
dilaton background \cite{kim} 
\begin{eqnarray}
  \label{ads2sols}
  (ds)^2 &=& -\left( -M +\frac{r^2}{\ell^2} \right) dt^2 +
  \left(-M + \frac{r^2}{\ell^2}  \right)^{-1} dr^2
\end{eqnarray}
where the horizon is located at $r_{H} = \sqrt{M}\ell$.
The auxiliary field in Eq.(\ref{auxiliary})  
is exactly solved 
as 
\begin{equation}
  \label{solution2}
  \Psi(r,t) = \frac{a \ell}{2\sqrt{M}} \ln \left
  ( 1-\frac{2\sqrt{M}\ell }{r+\sqrt{M}\ell} \right) - 
    \ln\left(-M +\left(\frac{r}{\ell}\right)^2 \right) + b + ct
\end{equation}
on the background (\ref{ads2sols}).
The integration constants $a,b,c$ will be determined by
some boundary conditions. 
By using Eqs.(\ref{ads2sols}) and (\ref{solution2}),
the explicit expression of the stress-energy tensors 
(\ref{stresstensor})
in terms of
the Schwarzschild coordinate is given as
\begin{eqnarray}
  \label{stressrt2}
  <T_{tt}> &=& \frac{1}{96}\left( c^2 + a^2 + \frac{4r^2}{\ell^4}
  - \frac{8M}{\ell^2}\right),\nonumber \\
  <T_{rr}> &=& \frac{1}{96}\left(\frac{r^2}{\ell^2}-M\right)^{-2} 
       \left( c^2 + a^2 - \frac{4r^2}{\ell^4}\right),
  \nonumber \\
  <T_{rt}> &=& \frac{ac}{48}\left(\frac{r^2}{\ell^2}-M\right)^{-1},
\end{eqnarray}
and in the light-cone coordinate they are simply  
\begin{eqnarray}
  \label{stresspm2}
<T_{\pm \pm}>&=& \frac{1}{192}\left( (c \pm a)^2 - \frac{4M}{\ell^2}\right),\\
<T_{+-}> &=& \frac{1}{48\ell^2} \left( \frac{r^2}{\ell^2}-M \right).
\end{eqnarray}
Now we impose the Hartle-Hawking boundary condition on 
the stress-energy tensors.
The black hole embedded in the thermal bath satisfies
the equilibrium condition which is described as
$<T_{\pm \pm}> |_{r \rightarrow r_{H}} = 0$
together with $c=0$ \cite{isr,hh,bfs}, 
and then the parameters are fixed as $a=\pm \frac{2\sqrt{M}}{\ell}$ and $b=0$.
If we make the auxiliary field finite at
the horizon, we can choose $a=\frac{2\sqrt{M}}{\ell}$.
This boundary condition determines the behavior of
the stress-energy tensors which are expressed
as 
\begin{eqnarray}
  \label{rescghspm2}
  <T_{\pm \pm}>_{HH} &=&  0,  \\
   <T_{+-}>_{HH} &=&  \frac{1}{48\ell^2}\left( \frac{r^2}{\ell^2}-M\right).
\end{eqnarray}
As expected, from the relation, $T_{H} = \frac{1}{\pi}T_{--}$ \cite{cghs}, the Hartle-Hawking state gives the
vanishing temperature of $ T_{H} = 0$ for the massless case,
and this fact is compatible with the scattering analysis
given in Ref. \cite{kop}.

Let us now study the massive radiation from
the AdS$_2$ black hole. Unfortunately,
we do not have an exact form of the
effective action unlike the massless case and the
calculation of the stress-energy tensor of the massive
scalar is not straightforward, which is recently discussed in Ref. \cite{ss}. 
Instead, we want to study the
Hawking radiation for the massive scalar field 
following the scattering procedure which is successful for 
the massless field case in Refs.\cite{kop,ko}. 

Then the massive scalar field equation is  
\begin{equation}
  \label{equationmass}
  (\Box + m^2 )f(r,t) = 0,
\end{equation}
where $m$ is a mass parameter. 
It is written as a spatial equation by using the ansatz, $f(r,t)=R(r) e^{-i\omega t}$, 
which is given as
\begin{equation}
  \label{spatialeqn}
  (r^2 - r_{H}^2) \partial_{r}^2 R(r) + 2r\partial_{r} R(r) + \frac{1}{(r^2
  -r_{H}^2)}[\omega^2 \ell^4 + m^2\ell^2 (r^2 - r_{H}^2)] R(r) = 0.
\end{equation}
By use of the change of variable
$z=\frac{r-r_{H}}{r+r_{H}}$ ($0<z<1$), 
the field equation (\ref{spatialeqn}) is written as
\begin{equation}
  \label{zwaveeqn}
  z(1-z)\partial_{z}^2 R(z) + (1-z) \partial_{z}R(z) + \left
  [ \frac{\omega^2 \ell^4}{4 r_{H}^2} \left(\frac{1}{z} - 1\right)
   + \frac{m^2 \ell^2}{1-z} \right]R(z) =0.
\end{equation}
To eliminate two singularities at $z=0$ and $z=1$, we set $R(z) =
  z^{\alpha}(1-z)^{\beta}g(z)$ and then the
wave equation
becomes
\begin{eqnarray}
  \label{intereqn}
  z(1-z)\partial_{z}^2 g(z) &+& [1+2\alpha -(1+2\alpha+2\beta)z]
  \partial_{z}g(z)\nonumber \\ 
  &+&\left[\frac{1}{z}\left(\alpha^2 + \frac{\omega^2
  \ell^4}{4 r_{H}^2}\right) + \frac{1}{1-z}\left(\beta^2 -\beta +
  m^2\ell^2\right) -\beta\left (\beta +2\alpha \right)\right]g(z) =0.
\end{eqnarray}
Here we choose $\alpha^2=-\frac{\omega^2
  \ell^4}{4r_{H}^2}$ and $\beta = (1-\sqrt{1-4m^2\ell^2})/2$,
and then get the final form of the field equation,
\begin{equation}
  \label{finaleqn}
  z(1-z)\partial_{z}^2 g(z) + [1+2\alpha -(1+2\alpha+2\beta)z]
  \partial_{z}g(z) - \beta\left (\beta +2\alpha \right)g(z) =0.
\end{equation}
Actually, we have two roots satisfying
 $\beta^2-\beta +m^2\ell^2 = 0$, however, 
we take the negative sign because there exists massless
limit($\beta=0$).
Note that we assume 
our semiclassical approximation is valid only for 
the case that
the energy of our test field is properly small, and 
we need not consider the back
reaction of the geometry. 
The curvature of the AdS geometry is
proportional to $\frac{1}{\ell^2}$, and 
we assume $m^2$ is less than $\frac{1}{\ell^2}$, $\beta$ 
is approximately $m^2 \ell^2$.

The field equation (\ref{finaleqn}) can be solved as
\begin{eqnarray}
  \label{solgen}
  R^{\rm bulk}(z) &=& C_{\rm out} z^{\alpha}(1-z)^{\beta} F(\beta,\beta + 2\alpha,
  1+2\alpha, z)\nonumber \\ &+& C_{\rm in} z^{-\alpha}(1-z)^{\beta} F(\beta,
  \beta-2\alpha,1-2\alpha,z).
\end{eqnarray}
This hypergeometric solution 
is symmetric under the exchange of $\alpha \rightarrow -\alpha$, 
so we may take the positive value of $\alpha$ for convenience.
In the near horizon limit
, the solution (\ref{solgen})
is explicitly written as
\begin{equation}
  \label{nearsol}
  R_{\rm near}^{\rm bulk}(r) = C_{\rm out} e^{i\frac{\omega
  \ell^2}{2r_{H}}\ln\left(\frac{r-r_{H}}{r+r_{H}}\right)} +C_{\rm in} e^{-i\frac{\omega
  \ell^2}{2r_{H}}\ln\left(\frac{r-r_{H}}{r+r_{H}}\right)}.
\end{equation}
In Eq.(\ref{nearsol}), it is independent of $\beta$ and the
asymptotic form of the field solution 
is remarkably same with that of the massless scalar
field \cite{kop}. This fact is plausible in that the 
wave frequency is ultraviolet shift in the
near horizon limit and the mass term
can be negligible compared to the wave number
in the dispersion relation. So in this region, the
scalar field can be regard as a massless conformal
field. 

Next, we can find the
far region solution by using the $z\rightarrow 1-z$ transformation of the
hypergeometric function \cite{as},
\begin{eqnarray}
  \label{z1zhyper}
  F(a,b,c;z) &=& \frac{\Gamma(c)\Gamma(c-a-b)}{\Gamma(c-a)\Gamma(c-b)}
  F(a,b,a+b-c+1;1-z) \nonumber \\&+& (1-z)^{c-a-b}
  \frac{\Gamma(c)\Gamma(a+b-c)}{\Gamma(a)\Gamma(b)} F(c-a,c-b,c-a-b+1;1-z).
\end{eqnarray}
Note that 
our transformed 
solution has no massless limit $\beta=0$
because the hypergeometric function of the transformed solution is not
defined for $\beta =0$. Thus this solution is only valid for the
case of the massive scalar field and it is meaningless at this stage to take the massless
limit. If $r$ goes to infinity, from Eqs.(\ref{solgen}) and (\ref{z1zhyper}), the far solution can be
expanded as 
\begin{eqnarray}
  \label{farsol}
  R_{\rm far}^{\rm bulk}(r) &=&(2r_{H})^{1-\beta} \left[ C_{\rm out}
  \frac{\Gamma(1+2\alpha)\Gamma(2\beta-1)}{\Gamma(\beta +
  2\alpha)\Gamma(\beta)} + C_{\rm in}
  \frac{\Gamma(1-2\alpha)\Gamma(2\beta-1)}{\Gamma(\beta -
  2\alpha)\Gamma(\beta)}\right] \frac{1}{r^{1-\beta}} \nonumber \\
&+&(2r_{H})^{\beta} \left[ C_{\rm out}
  \frac{\Gamma(1+2\alpha)\Gamma(1-2\beta)}{\Gamma(1-\beta +
  2\alpha)\Gamma(1-\beta)} + C_{\rm in}
  \frac{\Gamma(1-2\alpha)\Gamma(1-2\beta)}{\Gamma(1-\beta -
  2\alpha)\Gamma(1-\beta)}\right] \frac{1}{r^{\beta}} \nonumber \\
&+& {\cal
  O}\left(\frac{1}{r^{2-\beta}}, \frac{1}{r^{\beta+1}}\right).
\end{eqnarray}
Note that this asymptotic solution was derived from the
exact bulk solution. 
At this stage, 
we should carefully consider the boundary geometry of this
AdS black hole because it is nontrivial in 
contrast to the asymptotically flat black hole. 
The background geometry of the usual black hole at
the asymptotically far region is happened to be 
that of the massless limit
of the black hole geometry.
So, in that case, the far region limit means 
the massless limit of the black hole geometry. 
In our model, however, this is not the case.
Therefore, we should take a boundary metric by
defining $M=0$ in Eq.(\ref{spatialeqn}). 
Then the equation of motion at the boundary 
is given as 
\begin{equation}
  \label{beqn}
  r^2 \partial_{r}^2 R(r) + 2r \partial_{r} R(r) + \frac{1}{r^2} \left
  ( \omega^2 \ell^4 + m^2 \ell^2 r^2 \right) R(r) = 0
\end{equation}
by setting $r_{H} = 0$ from Eq.(\ref{spatialeqn}).
This equation (\ref{beqn}) yields the solution 
of a linear combination
of the Bessel functions $J_{\lambda}$ and $J_{-\lambda}$, 
\begin{equation}
  \label{bsolution}
  R_{\rm boundary}(r) = \frac{1}{\sqrt{r}} \left( A_{\rm 1} J_{-\lambda}
  \left(\frac{\omega \ell^2}{r}\right) + A_{\rm 2} J_{\lambda}
  \left(\frac{\omega \ell^2}{r}\right)\right), 
\end{equation}
where $\lambda$ is 
defined as $\lambda \equiv \frac{1}{2} -\beta$. 
For a large $r$, 
we can rewrite this solution
(\ref{bsolution}) as
\begin{equation}
  \label{bsol}
  R_{\rm boundary}(r) = B_{1} \frac{1}{r^{\beta}} + B_{2} \frac{1}{r^{1-\beta}}.
\end{equation}
To decompose this boundary solution 
into ingoing and outgoing modes, we define our
coefficients $B_{1}$ and $B_{2}$ in Eq.(\ref{bsol}) in terms of
new coefficients
$B_{\rm in}$ and $B_{\rm out}$,
\begin{eqnarray}
  \label{defcoef}
  & &B_{\rm 1} \equiv B_{\rm in} + B_{\rm out},\nonumber \\
  & &B_{\rm 2} \equiv \frac{i}{\pi}(2r_{H})^{1-2\beta}(B_{\rm in} -
  B_{\rm out}).
\end{eqnarray}
Then the boundary solution 
can be written as
\begin{equation}
  \label{bsolinout}
  R_{\rm boundary}(r) = \frac{1}{r^\beta}\left[B_{\rm in}\left( 1+
  \frac{i}{\pi} \left(\frac{2r_{H}}{r}\right)^{1-2\beta}\right) +
  B_{\rm out}\left(1-\frac{i}{\pi}\left(\frac{2r_{H}}{r}\right)^{1-2\beta} \right)\right]. 
\end{equation}
Now we match Eq.(\ref{farsol})  with Eq.(\ref{bsolinout}) to
get the relations between far coefficients and boundary
coefficients. 
Then two relations between these coefficients are obtained as
\begin{eqnarray}
  \label{matchcoef}
B_{\rm in} &=& \frac{1}{2}\left[(2r_{H})^{\beta}\left(C_{\rm
      out}\frac{\Gamma(1+2\alpha)\Gamma(1-2\beta)}{\Gamma(1-\beta+2\alpha)\Gamma(1-\beta)}+
      C_{\rm
      in}\frac{\Gamma(1-2\alpha)\Gamma(2\beta-1)}{\Gamma(\beta-2\alpha)\Gamma(\beta)}\right)\right.\nonumber
      \\&-&i\left. \pi(2r_{H})^{\beta}\left(C_{\rm
      out}\frac{\Gamma(1+2\alpha)\Gamma(2\beta-1)}{\Gamma(\beta+2\alpha)\Gamma(\beta)}+C_{\rm
      in}\frac{\Gamma(1-2\alpha)\Gamma(2\beta-1)}{\Gamma(\beta-2\alpha)\Gamma(\beta)}\right)\right],
      \nonumber \\
B_{\rm out} &=& \frac{1}{2}\left[(2r_{H})^{\beta}\left(C_{\rm
      out}\frac{\Gamma(1+2\alpha)\Gamma(1-2\beta)}{\Gamma(1-\beta+2\alpha)\Gamma(1-\beta)}+
      C_{\rm
      in}\frac{\Gamma(1-2\alpha)\Gamma(2\beta-1)}{\Gamma(\beta-2\alpha)\Gamma(\beta)}\right)\right.\nonumber
      \\&+&i\left. \pi(2r_{H})^{\beta}\left(C_{\rm
      out}\frac{\Gamma(1+2\alpha)\Gamma(2\beta-1)}{\Gamma(\beta+2\alpha)\Gamma(\beta)}+C_{\rm
      in}\frac{\Gamma(1-2\alpha)\Gamma(2\beta-1)}{\Gamma(\beta-2\alpha)\Gamma(\beta)}\right)\right].
\end{eqnarray}
The ingoing mode(outgoing mode) at the boundary can be
represented as ingoing and outgoing modes at the bulk \cite{bd}.
>From the following definition of the flux,
\begin{equation}
  \label{deflux}
  F \equiv \frac{2\pi}{i} 
\left(\frac{r^2 - r_{H}^2}{\ell^2}\right)
\left[R^{*}(r) \partial_{r}R(r) - R(r) \partial_{r} R^{*}(r)\right]
\end{equation}
we explicitly calculate the radiation flux at the 
boundary as 
\begin{eqnarray}
  \label{fluxinout}
  F_{\rm boundary}^{\rm in} &=& - \frac{4(2r_{H})^{1-2\beta}}{\ell^2}
  \sqrt{1-4m^2 \ell^2}|B_{\rm in}|^2,\nonumber \\
  F_{\rm boundary}^{\rm out} &=& \frac{4(2r_{H})^{1-2\beta}}{\ell^2}
  \sqrt{1-4m^2 \ell^2}|B_{\rm out}|^2
\end{eqnarray}
where we imposed the appropriate boundary condition 
that there does not exist
the outgoing mode 
near the black hole horizon, i.e., $C_{\rm out}=0$ \cite{gl}. 
Then the reflection coefficient is represented by
the ratio of the ingoing and the outgoing amplitude,
\begin{eqnarray}
  \label{reflcoef}
  R&=&\left|\frac{F_{\rm boundary}^{\rm in}}{F_{\rm boundary}^{\rm
  out}}\right|\\
   &=&\left|\frac{B_{\rm in}}{B_{\rm out}}\right|^2
\end{eqnarray}
and this formal expression can be explicitly evaluated
in the small mass  
compared to the given
curvature scale of AdS geometry, which is plausible
approximation in that 
we have not considered the back reaction of
the geometry. 
The useful formulas for some expansions of 
the gamma functions are summarized as follows,
\begin{eqnarray}
  \label{gammaexp}
& &  \frac{\Gamma(1-2a)}{\Gamma(1-a)} = 1 + \gamma a +
  {\cal O}(a^2),\nonumber \\
& &  \frac{\Gamma(2a-1)}{\Gamma(a)} = -\frac{1}{2} + \left(-1 +
  \frac{\gamma}{2}\right)a + {\cal O}(a^2), \nonumber \\
& &  \frac{1}{\Gamma(1+a+ib)\Gamma(1+a-ib)} =
  \frac{1}{\Gamma(1+ib)\Gamma(1-ib)}\left[1+(\psi(1-ib) +
  \psi(1+ib))a + {\cal O}(a^2)\right], \nonumber \\
& &  \frac{1}{\Gamma(a+ib)\Gamma(a-ib)} =
  \frac{1}{\Gamma(ib)\Gamma(-ib)}\left[1-(\psi(-ib) +
  \psi(ib))a + {\cal O}(a^2)\right], \nonumber \\
& &  \psi(1\pm a) = -\gamma \pm \frac{\pi^2}{3} a -4 \zeta(3) a^2 +
  {\cal O}(a^3), \nonumber \\
& &  \psi(\pm a) = \mp \frac{1}{2 a} + \psi(1\pm a),
\end{eqnarray}
where $\gamma$ is an Euler's constant, $\psi(z)$ is a digamma function,
and $\zeta(3)$ is the Riemann zeta function  
 $ \zeta(3)=\sum_{k=1}^{\infty}k^{-3}=1.20205\cdots$.
Then the reflection coefficient is simply given as
\begin{equation}
  \label{reflcoeff}
  R=\frac{4({r_{H}}^2 +2\zeta(3)\omega^2 m^2
  \ell^6)+\pi^2\omega^2\ell^4(1+4m^2\ell^2)-4\pi r_{H} \omega\ell^2(1+2m^2\ell^2)}{4({r_{H}}^2 +2\zeta(3)\omega^2 m^2
  \ell^6)+\pi^2\omega^2\ell^4(1+4m^2\ell^2)+4\pi r_{H} \omega\ell^2(1+2m^2\ell^2)}.
\end{equation}
We can obtain the Hawking temperature from
the following relation,
\begin{eqnarray}
\label{fomul}
 <0|N|0> &=& \frac{R}{1-R}\nonumber \\&=& \frac{1}{e^{\frac{\omega}{T_{H}}}-1}
\end{eqnarray}
where $N$ is a number operator \cite{bd}.
The Hawking temperature is simply expressed as
\begin{equation}
  \label{hawk}
  T_{H} = -\frac{\omega}{\ln R}
\end{equation}
in terms of the reflection coefficient.
Note that we expand the reflection coefficient with
respect to the frequency $\omega$ and the mass $m$,
and $\ln R$ term in the denominator 
is proportional to $\omega$, 
which yields the Hawking temperature 
\begin{equation}
  \label{hawkexp}
  T_{H} = \frac{r_{H}}{2\pi\ell^2}\left[1 -
  2m^2\ell^2\left(1 + 
\frac{\pi^2 \ell^2}{24 r_{H}^2}\right)\right] + {\cal O}(m^3).
\end{equation}
It would be interesting to note that the massless limit does
not recover the well-known null temperature. 
As was shown in Ref. \cite{kim,kop}
there does not exist the massless radiation on AdS$_2$ background,
and naturally the corresponding Hawking temperature vanishes.
For the massless case, the
transformation rule of the hypergeometric function
(\ref{z1zhyper}) can not be used because it is singular and the 
transformation rule for the massless case is somewhat different.
In this massive case, the nonvanishing result is 
in fact unexpected. One might think that
the massive Hawking radiation does not appear since
the massless radiation does not occur on the AdS background.
However, as was discussed in Ref. \cite{ko}, even for the
massless case the Hawking radiation is possible 
if the massless scalar field couples to some background fields,
for example, the dilaton field. So in the present massive case,
the mass term seems to play a role of the constant
background field.   
 
The final point to be mentioned is that 
the asymptotic behaviors for the cases of 
the massless field, the dilaton coupled massless field, 
and the present massive field. 
In the near horizon, the asymptotic behaviors of the
three cases are coincident, which means that
the dilaton coupled massless field and the massive
field behave like a conformal type massless field.
At the infinite boundary, however, three cases show 
different aspects unlike those of the near the horizon. 
First, for the massless field case, 
the boundary solution is given as
\begin{equation}
  \label{confo}
  R^{\rm boundary}_{\rm massless}(r) = A_{\rm out} e^{-i\frac{\omega\ell^2}{r}} +
  A_{\rm in} e^{i\frac{\omega\ell^2}{r}}.
\end{equation}
If $r$ goes to the asymptotic infinity, the boundary solution becomes
a constant. Next, for the massless field with the dilaton background, 
the boundary solution becomes
\begin{equation}
  \label{dilatonbs}
  R^{\rm boundary}_{\rm dilaton}(r) = 
A_{\rm out} \left(1-i\frac{r_{H}^2}{\pi r^2}\right)+A_{\rm in} \left(1+i\frac{r_{H}^2}{\pi r^2}\right).
\end{equation}
Note that this solution is also constant at the boundary. 
For the massive field case, 
the boundary solution is given as
\begin{equation}
  \label{massbc}
  R^{\rm boundary}_{\rm massive}(r) = A_{\rm out}
  \frac{1}{r^\beta}\left(1-\frac{i}{\pi}\left(\frac{2r_{H}}{r}\right)^{1-2\beta}\right)
+ A_{\rm in}
  \frac{1}{r^\beta}\left(1+\frac{i}{\pi}\left(\frac{2r_{H}}{r}\right)^{1-2\beta}\right),
\end{equation}
where it vanishes at the infinity 
due to the effect of the field mass. 
Therefore
the massive field is confined within the infinite boundary.
On the other hand, the key ingredient 
of the null radiation for the massless case is due to
the asymptotic forms of the field solutions.
For the massless case, the boundary solution is the same with the
far region solution in the AdS bulk, 
which means that the modes mixing does not occur,
whereas both for the dilaton coupled and the massive case,
the boundary solution is drastically different of the
far region solution from the bulk so that
the bulk solution can be decomposed into the ingoing and the
outgoing modes at the boundary.
Therefore, this fact gives the nonvanishing Hawking temperature in the latter two cases.
We conclude that in two dimensions the massive Hawking radiation of
AdS black hole is possible 
since the mass term in the field equation
plays a role of a constant background field,
which is similar to the dilaton coupled case. 

\vspace{20mm}

{\bf Acknowledgments}\\
We would like to thank M. S. Yoon 
for discussions.   
The authors wish to acknowledge the financial
support of Korea Research Foundation made in 
the program year of
1997.

\end{document}